\documentclass[sigconf, nonacm]{acmart}
\usepackage{graphicx}
\usepackage{enumitem}
\newcommand{\mycomment}[1]{}

\AtBeginDocument{%
  }

\setcopyright{acmlicensed}
\copyrightyear{2018}
\acmYear{2018}
\acmDOI{XXXXXXX.XXXXXXX}

\acmConference[Conference acronym 'XX]{Make sure to enter the correct
  conference title from your rights confirmation emai}{June 03--05,
  2018}{Woodstock, NY}
\acmISBN{978-1-4503-XXXX-X/18/06}

\begin{document}

\title{Apprentice Tutor Builder: A Platform For Users to Create and Personalize Intelligent Tutors}

\author{Glen Smith}
\affiliation{%
  \institution{Georgia Institute of Technology}
  \streetaddress{North Ave NW}
  \city{Atlanta}
  \state{Georgia}
  \country{USA}
  \postcode{30332}
}
\email{glensmith@gatech.edu}

\author{Adit Gupta}
\affiliation{%
  \institution{Drexel University}
  \streetaddress{3230 Market Street}
  \city{Philadelphia}
  \country{USA}}
\email{adit.gupta@drexel.edu}

\author{Chris MacLellan}
\affiliation{%
  \institution{Georgia Institute of Technology}
  \streetaddress{North Ave NW}
  \city{Atlanta}
  \state{Georgia}
  \country{USA}
  \postcode{30332}
}
\email{cmaclell@gatech.edu}


\begin{abstract}
Intelligent tutoring systems (ITS) are effective for improving students’ learning outcomes. However, their development is often complex, time-consuming, and requires specialized programming and tutor design knowledge, thus hindering their widespread application and personalization. We present the Apprentice Tutor Builder (ATB), a platform that simplifies tutor creation and personalization. Instructors can utilize ATB’s drag-and-drop tool to build tutor interfaces. Instructors can then interactively train the tutors’ underlying AI agent to produce expert models that can solve problems. Training is achieved via using multiple interaction modalities including demonstrations, feedback, and user labels. We conducted a user study with 14 instructors to evaluate the effectiveness of ATB’s design with end users. We found that users enjoyed the flexibility of the interface builder and ease and speed of agent teaching, but often desired additional time-saving features. With these insights, we identified a set of design recommendations for our platform and others that utilize interactive AI agents for tutor creation and customization.
\end{abstract}




\maketitle

\section{Introduction}
Intelligent tutoring systems (ITSs) are an educational technology that provide students with course content, guided-practice problems, and performance feedback \cite{10.5555/1435351.1435353}. This kind of tutored support has been shown to be effective at improving students' learning outcomes \cite{ma2014intelligent}. ITSs can track a student's knowledge acquisition and progression and select personalized problem sequences to optimize the student's learning. These tutors can be deployed to many students at once, offering a scalable solution to issues of high class volumes and needs for supplemental learning to name a few.

However, several limitations prohibit the wide-spread use and applicability of ITSs. One limitation is that tutor development often requires specialized programming and tutor design knowledge \cite{murray2003overview}. In particular, expert models that can evaluate learners as they solve problems must be authored for each tutor, an often cumbersome and tedious task. This prohibits non-technical users from creating and customizing their own tutors \cite{ctat_tutor} and limits tutor creation to those with the required knowledge, such as academic researchers and UI and software developers. Another limitation is that development is time-consuming \cite{murray2003overview}. Previous work estimates that for well-established tutor authoring methods such as model-tracing it takes approximately 200-300 hours of tutor development time to produce just 1 hour of instruction time \cite{Weitekamp2020}. Other approaches such as example-tracing \cite{aleven2009example} and the Apprentice Learner framework \cite{MacLellan2020} have reduced authoring time even further, but there is still work to be done to make authoring accessible to everyday users. These issues prohibit the ability to scale tutor production to meet the individual needs of every classroom, and as a consequence, tutors are often developed with one-size-fits-all designs. These tutors are generally designed to be as widely usable as possible and are unable to be customized to align with users' specific needs.


To address these challenges, we present the Apprentice Tutor Builder, a tutor authoring platform that allows users to create and customize tutors. ATB lets users create tutor interfaces in a row-column layout using a drag-and-drop tool. Additionally, users author the underlying AI expert models interactively via demonstrations, correctness feedback, and user-provided labels. To evaluate the usability of our system, we conducted a user study with 14 instructors where each participant built tutors and trained expert models for two tasks, fraction arithmetic (addition and multiplication) and Square 25, a procedure for squaring any number that ends in 5. From this study, we evaluated the usability of ATB to inform a set of design recommendations for similar systems and make the following key claims:

\begin{itemize}
  \item The row-column layout approach to the interface builder is usable by teachers without any specialized training.
  \item Our interactive approach lets teachers author expert models via demonstrations, correctness feedback, and dialog.
  \item Teachers can see the value and potential for this technology and express a willingness to engage in tutor creation and customization.
\end{itemize}

Interactive machine-learning based tutor authoring tools have been tested within the lab \cite{maclellan2022domain,Weitekamp2020}, but to the best of our knowledge this is the first paper that has tested such approaches directly with teachers, demonstrating their potential with target end users. We believe demonstrating these claims is a first step towards showing that ATB has the potential to facilitate customized tutor creation and deployment at scale.

\mycomment{In the following section, we present related work. In section 3, we present a full system description. Section 4 presents the user study design and results. We discuss these results in section 5 and conclude with section 6 where we discuss future work. }

\section{Related Works}
Aleven, et. al. (2019) describes a six step process for building example-tracing intelligent tutors \cite{Aleven2009}. While these steps are specific to example-tracing, some of these concepts are general to building any tutor. First, the tutor author investigates student thinking and learning in the given task domain using cognitive task analysis \cite{lovett1998cognitive}. Based on insights from this analysis, the author designs a tutor interface for a specific problem type. Authors break down complex problems into simpler steps. Finally, the author creates an expert model to power the tutor. Recent authoring tools let authors use programming by demonstration to achieve this \cite{matsuda2010learning, exampletracing}. 

We are particularly interested in a specific type of intelligent intelligent tutor, known as a cognitive tutor \cite{anderson1995cognitive}. Cognitive tutors often include capabilities for individualized problem selection, on-demand hints, and real-time correctness feedback \cite{anderson1995cognitive}. These capabilities, while difficult to program, are correlated with successful student learning outcomes \cite{koedinger_1997}. Building cognitive tutors is not trivial, requiring extensive time and expertise to program an interface for each tutor type and production rules for the cognitive model. These high costs have led researchers to investigate how we might effectively and efficiently build intelligent tutors with less need for specialized skills. 

Intelligent tutor authoring tools aim to achieve this by providing a mechanism for building tutor interfaces and authoring expert models without programming. Recent work on ITS uses programming by demonstration techniques to let users author expert models by providing correctness feedback and demonstrations to build expert models. This approach lets users achieve model completeness in less time than exiting authoring approaches such as CTAT’s example tracing or SimStudent. There have been a plethora of authoring systems that reduce the time to build intelligent tutors. However, a key gap has been that these systems rarely utilize AI, and those that do have not bee tested with real end-users \cite{Weitekamp2020, MacLellan2022}.

\subsection{Building the Tutor Interface}
Building a tutor interface is an important part of the intelligent tutor, however, there is little prior research which reviews and compares the different types of tutor interfaces. Tutor authoring tools which come with an interface builder can be bucketed into a few categories including drag-and-drop systems, templates interface builders, and more recently, chat based. While this is not an exhaustive review of all types of interfaces, these three are the most common amongst tutor builders. 

Authoring tools such as CTAT, utilizes a drag-and-drop interface builder which allows users to drag specific components into a fixed location pane to create interfaces. While drag-and-drop interfaces are intuitive, a key issue with drag and drop functionality and fixed location components within CTAT is that the interfaces are inflexible to slight variations of the problem type.

Other systems like ASSISTments use a template based approach to create tutor interfaces \cite{heffernan2006assistment}. ASSISTments allows users to utilize text boxes, multiple choice, and drop-down components, through a web-based interface design experience. Template-Based Tools, rely on predefined templates, making them user-friendly but less adaptable to diverse domains or question types. While they can emulate problem-solving steps for specific scenarios, they limit users to the question structure within the template, often accommodating only a single question \cite{assistment_tutor}. 

In recent years, generative artificial intelligence (GAI) tools are diffusing rapidly. Educational institutions and EdTech companies are swiftly formulating strategies to harness the power of artificial intelligence (AI), particularly focusing on the effective utilization of GAI tools. This includes the incorporation of large language models (LLMs), such as ChatGPT. A key motivation for the use of GAI within educational institutions, is that their chat experience aims to increase engagement, which is correlated with retention \cite{kshetri2023economics}. Studies indicate that heightened student engagement is vital for achieving positive academic results and preventing adverse outcomes, such as dropout \cite{fredricks2004school}. However, LLMs have some severe weaknesses, including the ability to make mistakes, "hallucinate" false information, and generate biased outputs \cite{alkaissi2023artificial}.

\subsection{Authoring Expert Models}
Building a cognitive model for a specific tutor interface can be challenging, given the complexity of the problem type. Constructing a cognitive model through programming can be laborious, necessitating 200-300 hours of programming for each hour of tutor instruction. Tutor authoring tools aim to reduce the time required to construct tutors by employing techniques such as programming by demonstration to facilitate the building of expert models. CTAT utilizes a process called example-tracing, which enables users to create ITS content and expert models without the necessity of programming. Example-tracing simplifies the process by allowing individuals to demonstrate problems, with the correct actions being recorded in a behavior graph. This graph encapsulates knowledge in the form of a node-link structure. However, a limitation of this method is that it requires the demonstration of each possible solution to a problem, potentially leading to an incomplete model if the user is not exhaustive in their demonstrations. CTAT attempts to mitigate this with mass production techniques, yet this approach remains time-consuming and necessitates the use of Excel spreadsheets, presenting a substantial learning curve for novice users.

Recent advancements in CTAT have introduced the use of interactive machine learning, alongside programming by demonstration, to construct expert models more efficiently \cite{maclellan2014authoring}. This innovation lets authors leverage demonstrations and correctness feedback to develop comprehensive expert models that generalize beyond the initial examples used to train the system. While this approach to interactive machine learning represents a novel application within the realm of tutor authoring tools, an existing gap, which we aim to bridge, is the deployment of authoring tools that utilize interactive AI techniques to actual users, such as teachers.

\begin{figure*}[!t]
  \includegraphics[width=\linewidth]{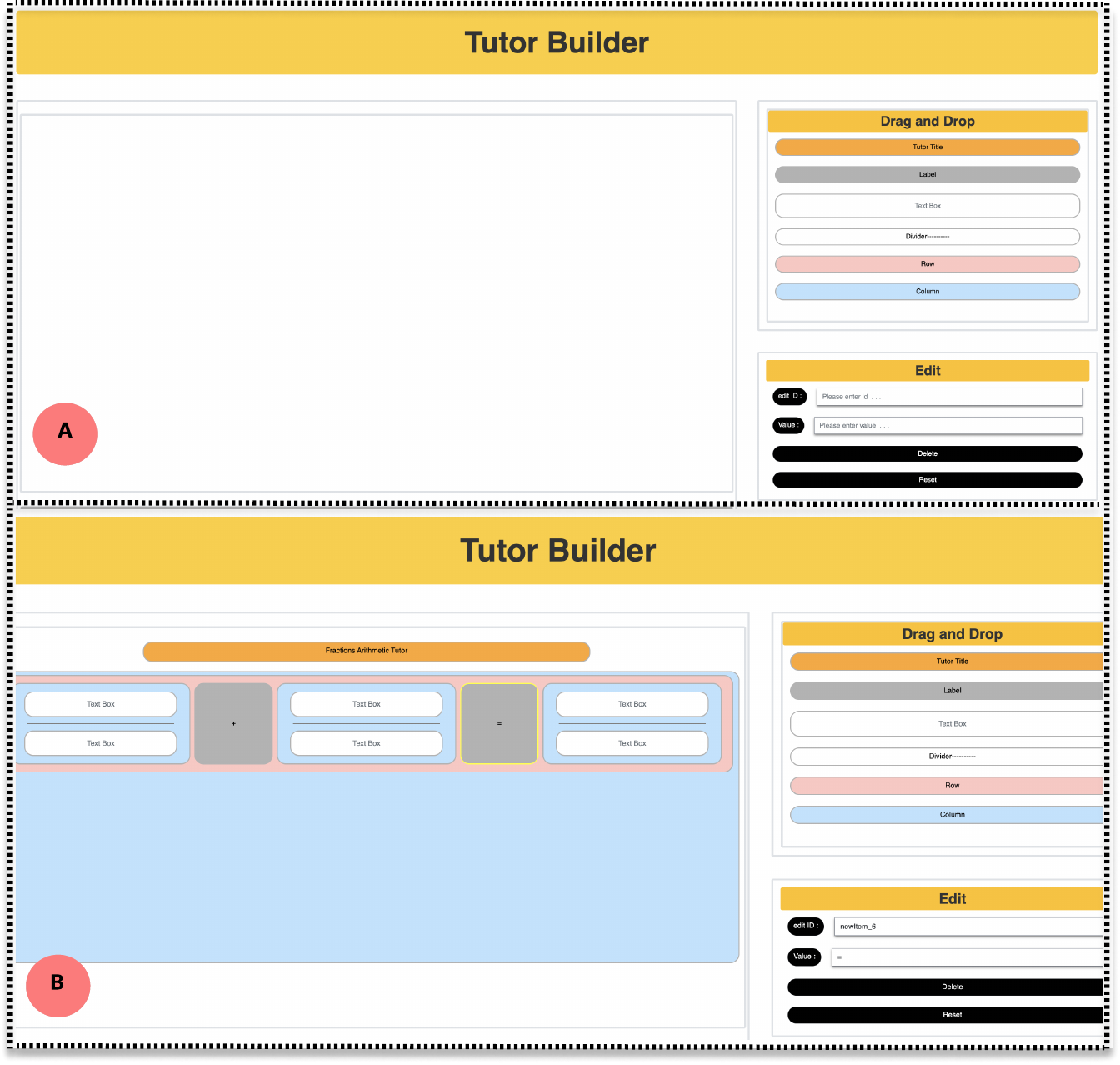}
  \caption{Apprentice Tutor Builder: (a) the interface builder starts from a blank canvas, and (b) users can drag and drop components from the right component box to the draggable pane area. }
  \Description{}
  \label{fig:drag-drop}
\end{figure*}

\section{System Description}
To help inform the design of Apprentice Tutor Builder (ATB), we conducted preliminary needs finding with teachers at a state technical college. We deployed four mathematics tutors to College Algebra class sections and following the deployment, we conducted several focus group sessions to collect feedback on teacher and student experiences on working with the system. We found that among instructors there was a significant preference for utilizing tutors whose content aligns with their course syllabus. Furthermore, every instructor has a unique teaching style, which can pose a challenge when attempting to align the capabilities of an intelligent tutor with their individual teaching methods. 

In this section, we describe the components of ATB, which take inspiration from CTAT \cite{ctat_tutor}, Apprentice tutors \cite{MacLellan2022}, and ASSISTment tutors \cite{assistment_tutor}. We extend the ideas and theories of these systems to create a novel approach of building intelligent tutors. This system allows users to author tutors and tutor content via two components: the tutor builder and apprentice agent training module.

\subsection{Tutor Builder}

The tutor builder draws inspiration from intuitive drag-and-drop builders, such as CTAT \cite{ctat_tutor}, and form-based builders like ASSISTments tutor \cite{assistment_tutor}. Our primary objective in developing the tutor builder was to empower non-technical users to create pedagogically robust tutors. We facilitate this through an intuitive user interface that offers users the flexibility to design diverse and stylistically varied interfaces. 

Figure \ref{fig:drag-drop} shows the empty builder interface (on the top) juxtaposed with an authored tutor interface (on the bottom). The interface builder utilizes the spring-and-struts layout approach \cite{scoditti2009new}, which lets users design their tutors in nested rows and columns. Users can drag elements such as rows, columns, input fields and labels into the designated pane on the left (component 1). Additionally, users can nest rows inside of columns and vice versa to create structured designs that auto-align interface elements. The row-column approach creates inherent relationships between interface elements (such as "a is contained in b") which the underlying expert model can use when learning how to solve problems (See section 3.2). To our knowledge, the integration of a row-column feature is a novel approach for tutor interface builders. 

\subsection{Apprentice Agent Training}
Underlying each tutor is an apprentice agent that users must train. In this section, we describe the process by which users interact with an agent to teach it how to solve tutor problems. We follow this with a description of the agent architecture in terms of its knowledge representation, learning, and performance components. 

\subsubsection{Authoring Experience}
During agent training, the user authors the expert model by providing problems and engaging with the agent via a series of interactions. First, the user initializes the problem by supplying the initial values needed to solve the problem. For example, to teach an agent how to add or multiply two numbers, a user might initialize the problem by entering the two operands and the operator into the interface. Once the problem is set up, the user clicks "Start Problem" which begins agent training. 

\begin{figure}[!t]
  \includegraphics[width=0.4\linewidth]{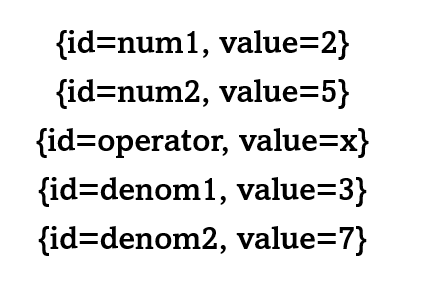}
  \caption{A sample working memory for the fraction arithmetic task. The \emph{facts} of the tutor consist of the field names and their values.}
  \Description{An example of fraction arithmetic tutor state/working memory}
  \label{fig:working-memory}
\end{figure}
\begin{figure}[!t]
  \includegraphics[width=\linewidth]{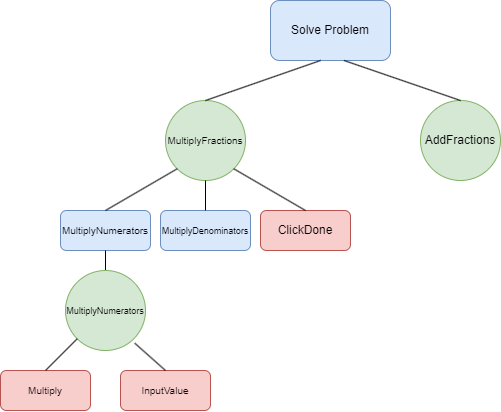}
  \caption{A sample HTN decomposition for fraction multiplication. The blue squares represent tasks, the green circles are methods, and the red squares are operators.}
  \Description{An example of fraction arithmetic HTN model}
  \label{fig:htn}
\end{figure}

The current state of the tutor---containing all values and the fields in which they appear---is sent to the agent\ref{fig:working-memory}.  shows how the tutor state is represented to the agent. At first, the agent does not have knowledge of how to solve the problem and asks for a demonstration. The user provides a demonstration by completing the next step directly in the interface. Next, the agent asks for a task label for the demonstration. These labels default to the unique identifier of the field but providing a label lets the agent to provide more meaningful and interpretable explanations later. Another benefit is that we can use these user-supplied task labels when evaluating student performance data to engage in labeling of knowledge components for knowledge tracing.

Given the demonstration and task label, the agent updates its knowledge. If the agent does not have sufficient knowledge to attempt the next step, it asks for another demonstration and the process continues. If the agent is able to attempt the next step, it will automatically complete it by filling in the respective field. The agent then asks "Did I take the correct action?" and requests yes/no feedback from the user. To aid the user in understanding its reasoning, the agent provides a text-based explanation of the steps it is taking and highlights which fields it is using to generate the result. If the user clicks "yes", the agent will move on to attempt the next step. If the user clicks "no", the agent removes its response and if it is able, attempts the step again, applying a different response. If no response is deemed correct by the user, the agent requests a demonstration. When the problem is complete, the user clicks "Done" and the interface is reset to let the user continue training by providing additional problems. This process continues until the user is satisfied that the agent can solve the problems fully and correctly.

\subsubsection{Knowledge Representation}
The apprentice agent utilizes two main knowledge structures: a Hierarchical Task Network (HTN) \cite{erol1994umcp} used to perform planning when solving problems and a \textit{working memory} which consists of facts about the tutor interface. HTNs are a knowledge framework for representing and solving complex problems by breaking down tasks into hierarchically organized subtasks \cite{erol1994umcp} and ultimately into operators it can execute. HTNs consist of a set of tasks to be completed, a set of methods that achieve those tasks, and a set of operators that represent the agent’s fundamental knowledge. The “network” is represented as an AND-OR tree, where to solve a task, only one valid method of that task needs to be chosen (OR), and each method contains a set of subtasks which must all be completed to successfully complete that method (AND).

In ATB, tasks are the problem step labels supplied by the user during training. Methods represent the various ways to solve a problem. For example, on a fractions arithmetic task, there may exist a high-level task "Solve Problem", for which there may be two methods \textit{AddFractions} and \textit{MultiplyFractions}. To determine when a method is applicable, methods contain conditions which must match the current tutor state. Continuing our example, one might apply \textit{AddFractions} when there exists a plus sign (+) in the tutor and one might apply \textit{MultiplyFractions} when there exists a multiplication symbol ($cross$) in the tutor. These conditions are acquired and updated during the HTN learning process and in general there may exist many conditions on a method beyond those illustrated here \ref{fig:htn}. 

Additionally, each agent is provided a set of primitive operators which can be placed in two categories: (1) operators that manipulate the working memory. These operators are analogous to "mental steps" such as the ability to add/multiply/subtract/divide two numbers, and (2) operators that take an action in the tutor interface such as inputting a value into a field or clicking the "Done" button. A special set of operators called \emph{relations} are also supplied. These operators define relationships over the values in the tutor, such as \emph{less-than} (a < b) or \emph{equals} (a == b) in a process called \textit{relational inference}. Relations are an important discriminator in the case of condition learning for methods. For example, given that there is a different process for adding two fractions with same denominators vs. with different denominators, we might further decompose \textit{AddFractions} into methods \textit{AddSameDenom} and \textit{AddDiffDenom} with an important condition for \textit{AddSameDenom} being that denominator 1 \textit{equals} denominator 2 (d1 == d2).

Finally, the working memory is a combination of data extracted from the tutor interface and information derived from that data such as relations between objects or values in the interface. Figure \ref{fig:working-memory} shows an example of a working memory state for a fraction arithmetic task.  

\subsubsection{Performance Component} 
To solve problems, the agent is given a task along with the tutor state and utilizes its HTN to plan and generate a response to the tutor interface. Here, tasks correspond to labels users provided during learning (See Learning Component). When a task and tutor state is observed, the agent first converts the state into \emph{facts} (figure \ref{fig:working-memory}), working memory objects representing fields in the interface, and uses a rete network \cite{forgy1989rete} to perform relational inference and produce new facts that are added to the working memory.  Next, the rete network is used to match the current facts in the working memory to an applicable method that satisfies the current task. If a matching method is found, the agent will select that method and then subsequently attempt to solve all subtasks by finding matching methods that solve each subtask. Once an operator is encountered in the decomposition, it is executed and its effects are added to the working memory. If the operator produces a value for the interface, this value is returned as the solution to the next step in the problem. At any point during planning, if no matching method is found for a task, the agent returns False, indicating that it cannot produce a solution given the current state.

To make this process more concrete, figure \ref{fig:htn} shows a decomposition of a fraction multiplication problem. figure \ref{fig:working-memory} shows the initial working memory and  figure\ref{fig:htn} shows the HTN decomposition. First, the \textit{MultiplyFractions} method is selected because the state matches the condition requiring a multiplication symbol to be present in the state. This method decomposes into three subtasks, \textit{MultiplyNumerators} , \textit{MultiplyDenominators}, and \textit{ClickDone}, which must all be solved to complete the parent method. Each subtask has one method of the same name that decomposes finally into two operators, \textit{Multiply} and \textit{InputValue}, where the former multiplies the values of the numerators and the latter takes the result and generates a return value to the interface. Finally,  the HTN planner solves the \textit{ClickDone} subtask using a special operator for clicking the done button and ending a problem.

\subsubsection{Learning Component}
The agent learns from demonstrations, feedback, and labels interactively provided by the user. Learning new methods consists of two parts: generating the subtask decompositions and iteratively adjusting their conditions. We discuss each in turn.

Initially the agent lacks any methods for solving tasks and only contains pre-authored operators. When the tutor state is observed, planning occurs similar to the performance component, however when no matching method is found, the agent requests a demonstration from the user. Upon receiving a demonstration, the agent must then \textit{explain} the observed value. To generate an explanation, the rete network iteratively matches facts to each applicable operator and executes them to generate changes to the working memory. This continues until either the rule trace provides an explanation for the demonstration, or the maximum depth parameter is reached. If the agent finds an explanation, it constructs a new method where its subtasks correspond to the sequence of operators found to generate the user’s demonstration. Alternatively, if the maximum depth parameter is reached, the agent simply creates a new method whose output is the given demonstration. This new method  functions as a \textit{memorized operator}. It only applies to the current tutor state and always outputs the same value\footnote{An explanation might not be found because the demonstrated value requires more levels of search, or "mental steps", than the allowed max search depth (usually 2 for performance purposes) or the user may have supplied the incorrect value and there is no possible way to reproduce the value given the current tutor state and the agent's operators.}. In both cases, the conditions of the method are the conjunction of facts in the working memory and the method is then categorized in the HTN under the task name provided by the user.

\begin{figure*}[!t]
  \includegraphics[width=\linewidth]{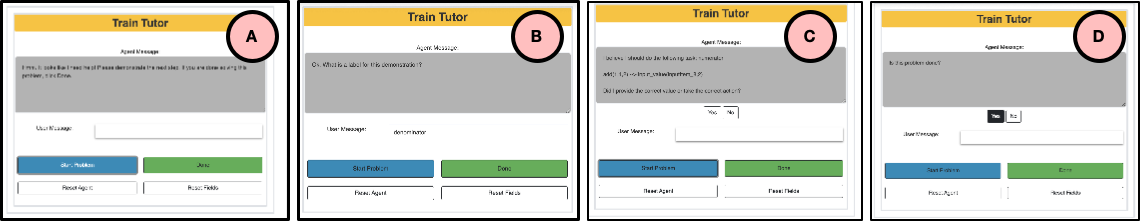}
  \caption{Interactive dialogs for training the expert model: (A) the system asking the user to provide a demonstration, (B) The agent requesting a label for a previously provided demosntration, (C) the agent requesting feedback on the correctness after taking a step, and (D) the agent asking if the problem has been correctly completed.}
  \Description{An example of agent training interactions}
  \label{fig:apprentice_tutor_builder}
\end{figure*}

Although demonstrations help the agent learn new methods, the agent also utilizes feedback from the user to determine \textit{when} to apply learned methods. During training, when the agent attempts to solve the next step of a problem, it prompts the user: "Did I take the correct action?" and requests a Yes/No response. To aid the user in providing accurate feedback, the agent also provides a text-based explanation of its solutions as well as highlights the fields used in its computation (see Figure~\ref{fig:apprentice_tutor_builder}, frame C). If the user selects "Yes", the agent simply continues to attempt the next step. If the user selects "No", the agent continues planning to locate another applicable method to solve the next step, if one exists. If not, the agent requests a demonstration for that step and engages in the explanation process described in the previous paragraph. When a new method is created, it is compared against all other methods for the current problem step (or task). If any two methods are found to have the same subtask decomposition, their conditions can be generalized to create a single method. Conditions are generalized using least-general generalization \cite{plotkin1970note}, employing the anti-unification algorithm to variablize values that differ from one set of conditions to another, leaving all others as constant. Updating the conditions in this way ensure that the method is applicable for future tutor states similar to those the method was created for or was previously correctly applied to. As will be seen in the study results section, methods and their conditions can converge after a small number of demonstrations and feedback from the user. 


\section{User Studies and Evaluation}
We performed a user study to evaluate the effectiveness and usability of ATB. Participants were asked to complete two tasks. Each task consisted of two components: (1) using the tutor interface builder to create a tutor and (2) authoring an expert model for the tutor  by interactively providing demonstrations, feedback, and labels to the AI agent. We followed these activities with a semi-structured interview designed to gauge user experience with and reactions to the system as well as to help inform future design choices and design recommendations for tutor authoring systems like ATB.

\subsection{Participants}
For this study, we recruited 14 participants (ages 18-54). We targeted participants with some degree of teaching experience, with only one participant having none. We also asked participants to gauge their experience levels in three other categories: Programming/Coding, Mathematics, and Artificial Intelligence. Additionally, of those with teaching experience, we recorded which levels they have taught, ranging from Primary School to Post-Secondary/Trade school and Government/Private Industry. These breakdowns are reflected in Figure \ref{fig:experience_level_likert_matrix}.

\subsection{Methods}
\begin{figure}[!t]
  \includegraphics[width=\linewidth]{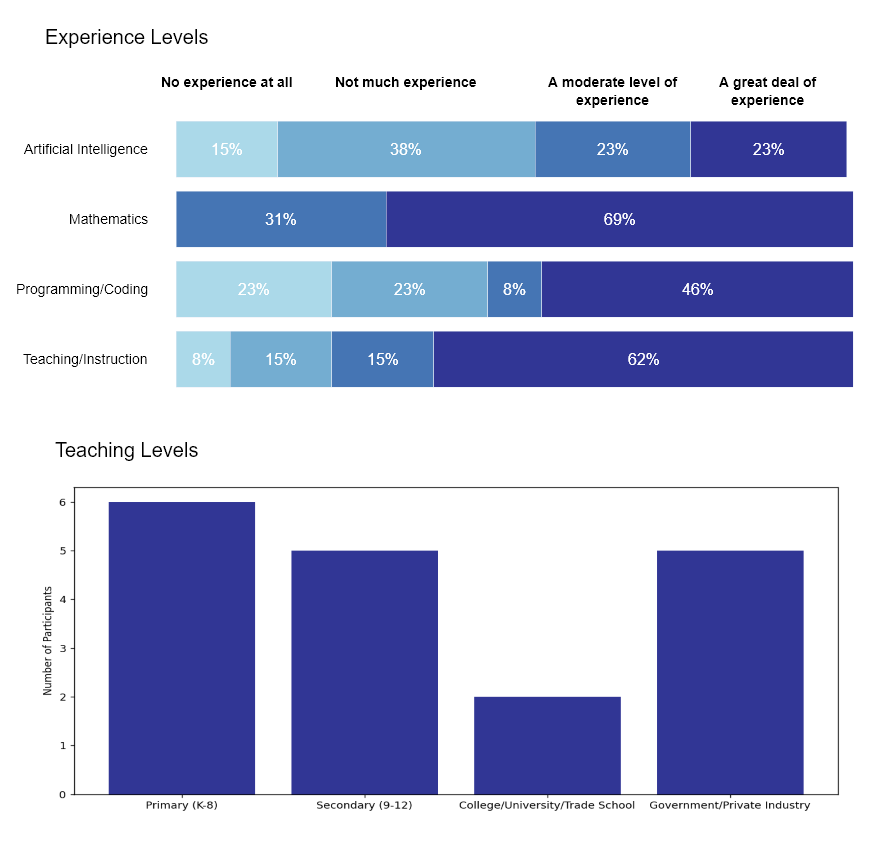}
  \caption{Experience levels of participants in four categories (1) artificial intelligence, (2) mathematics, (3) programming/coding, and (4) teaching (top) and instruction levels taught (bottom)}
  \Description{Experience levels of Participants}
  \label{fig:experience_level_likert_matrix}
\end{figure}

Before each session, we obtained consent from each participant and afterward provided a compensation of \$15/hour rounded up to the nearest half hour. At the beginning of the session, we collected demographic data such as age group, teaching, programming, math, and AI experience, as well as instruction levels taught  (see Figure~\ref{fig:experience_level_likert_matrix}). Next, we provided a 3-minute video tutorial. The video demonstrated both tutor interface building using the drag-and-drop tool and expert model authoring using the agent training interface and side panel. We allowed the participant to ask any questions at this stage to clarify the overall session structure as well as how to use the various components of the system. 

In the first task, participants were asked to use the tutor builder to create a \textit{Fractions Arithmetic} tutor (see Figure~\ref{fig:apprentice_tutor_builder}, part B). We then asked participants to teach the agent both fraction multiplication and fraction addition with same denominators. Users were asked to initialize each problem by providing four values for the two fraction operands as well as a plus (+) or multiplication (x) sign as the operator. Once the problem was initialized, users would click "Start Problem" to begin agent training.

In the second task, we asked participants to create a tutor for the \textit{Square 25} procedure. Square 25 is a 4-step process to easily square any number that ends in 5. The tutor contained the following fields: 
\begin{itemize}
  \item \textbf{Initial Value}: This is the value that the user wants to square
  \item \textbf{First Part}: This value is everything to the left of the 5 in the ones place
  \item \textbf{Add One}: Add 1 to the value of "First Part"
  \item \textbf{Multiply}: Multiply the resulting values of the "First Part" and "Add One" steps
  \item \textbf{Append 25}: Concatenate 25 to the end of the "Multiply" value
\end{itemize}
Similar to the first task, participants were asked to initialize problems by completing the "Initial Value" field and clicking "Start Problem".

For both tasks, participants were allowed to initialize problems however they chose. For the Fraction Arithmetic tutor, participants were allowed to teach fraction addition and multiplication in any order. This variation among participants meant that the resulting expert models would differ from user to user as it would in real-world scenarios. Allowing users to specify their own problems allowed us to evaluate the resulting expert models and inform future development opportunities for agent training. We also left it up to the participant to decide when each expert model was sufficiently trained. Knowing when the agent has sufficiently learned a task is a hard problem \cite{maclellan2022domain}, so evaluation of learned models should help us determine how we might improve the authoring experience. 

Once participants completed both tasks, we conducted a semi-structured interview where we gauged perceived ease and intuitiveness of the interface builder and agent training and overall usability of ATB. We video and audio recorded all tutor authoring sessions and logged all user-system interactions using an integrated system logger. Finally, we conducted two types of analyses on the participant data. First, we performed a quantitative analysis of user logs to determine how different user characteristics might correlate with usability. Next, we used \textit{affinity diagramming} \cite{lucero2015} to conduct a thematic analysis and derive concepts from the open-ended responses collected during the post-study interview and other comments noted throughout the session. We discovered a set of overarching themes that provide insight into the needs of ATB users.

\begin{figure*}[!t]
  \includegraphics[width=\linewidth]{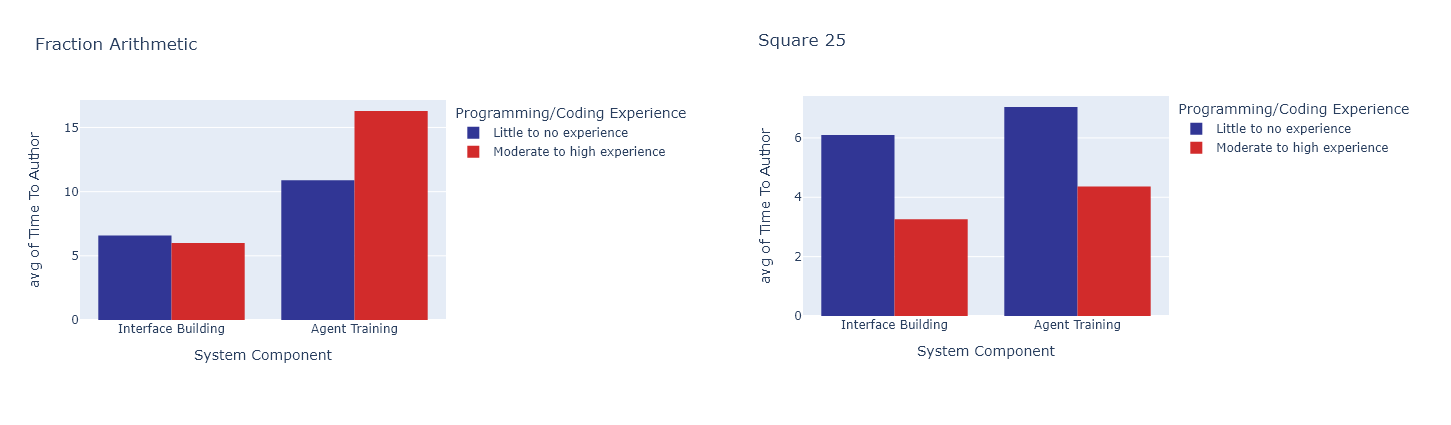}
  \caption{Average authoring time of the interface and agent training components of ATB for each the Fraction Arithmetic (left) and the Square 25 tutors (right). Averages are broken down by programming experience.}
  \label{fig:time_to_author}
\end{figure*}

\subsection{Results}
All 14 participants were able to fully recreate both tutors using the tutor builder and 13 of out 14 were able to author expert models that could solve both tasks. We present results in the following subsections, broken up by analysis technique.

\subsubsection{User Log Analysis}
Informed by prior work \cite{li2019pumice}, we grouped participants into two groups based on reported programming experience levels, those with little to no experience and those with moderate to high experience, or the "Low" and "High" group, respectively. Because the primary audience for our system is users with less technical expertise, determining the relationship between experience level and usage patterns allows us to understand how expertise in these areas relates to the usability our tool. Figure~\ref{fig:time_to_author} shows the average authoring time for both the interface builder and agent training components of ATB for each tutoring task. Overall authoring time is higher for the Fractions Arithmetic (FR) task (Avg: 11.85 min) compared to the Square 25 (S25) task (Avg: 7.12 min)\footnote{Time to author the fractions arithmetic task is expected to be higher since the task requires teaching the agent multiple problem types.}. For the FR tutor, those in the Low group spent slightly more time building the interface than the High group (Avg: 6.2 min vs. 5.5 min), but spent significantly less time training the expert model than the High group (Avg. 10.9 min vs. 16 min).  For the S25 tutor, the Low group spent a similar amount of time building the interface as the FR tutor while the High group spent almost half the time of the Low group on interface building (Avg: 6.1 min vs. 3.2 min). Finally, in contrast to the FR tutor, the Low group spent more time training the expert model than the High group (Avg: 7 min vs. 4.2 min). 

To explore how well the expert models were trained, we tested the models for each tutor to measure model correctness. We define model correctness (or accuracy) as the successful solve rate of 20 randomly generated problems for the FR expert model, 10 for multiplication and 10 for addition, and 10 randomly generated problems for the S25 expert model. All expert models were tested with the same 30 problems. All participants in the High group (moderate to high exp.) were able to author a model with 100\% model correctness for both tasks. However, for one participant in the Low group, the model correctness rate decreased to 20\% when solving fraction multiplication problems in the FR tutor. All participants in both groups achieved 100\% model correctness for the S25 tutor. 

Next, we examined the number of  \emph{validation problems} completed by each group. We define a validation problem as any additional problem used to train the agent after it has demonstrated the ability to fully solve a problem. Thus, if an agent fully solves a problem, and the user provides two more additional problems for the agent to solve, the number of validation problems is 2. Examining the number of validation problems provided by users allows us to determine how users' prior experience correlates with their ability to author a more correct model. For the Low group, the average number of validation problems increased from 0.33 to 1.0 when going from the FR tutor to the S25 tutor. In contrast, this number decreased from 2.25 to 0.15 for the High group. We hypothesize that those with more programming experience may initially be more cautious when training, but may more quickly develop an intuition for when the agent is sufficiently trained. 

\subsubsection{Affinity Diagramming}
During our thematic analysis of user comments, eight main themes emerged. We will discuss each in this section.

\textbf{\textit{Teacher Benefits}} Many participants praised the straightforward and intuitive nature of ATB's tutor authoring process and stated key benefits to teachers such as time savings. P7 said: "...saves a lot of time. And we'' be able to accomplish more [in] less time because the course is really big. Sometimes we run short of time because of that". Similarly P4, 6, and 10 all commented on how quick agent training was with one stating: "It was pretty easy...the agent picked up right away after like three problems". Additionally, participants expressed a desire to integrate ATB into their own classroom environments. P3 said: \emph{I find it very useful, especially [working] in the school, and with the students, it's gonna be very, very helpful to teach them}. They follow this with: \emph{If it's in my class, I'll definitely use it}.

\textbf{\textit{Student Benefits}} In addition to teacher benefits, participants also identified how students might benefit from ATB. P7 said that the system would be "more productive for the students" and P3 added that the ability for students to practice different questions would be "very useful" for them. P3 also noted that ATB can increase the instructional presence of the teacher, allowing for additional help and practice when the teacher is unavailable to the students. Along another angle, P18 mentioned that teachers could "allow students to build their own tutors [and] teach them or maybe turn [the authoring process] into a game". 

\textbf{\textit{Interface Builder: User Control}} Some participants noted that they would have liked more flexibility in the interface building component. For example, participants requested more control over the ability to freely align and order elements on the screen. P6 said: \emph{It would be great if I could space things out how [I want], like the multiplication or addition symbol floating at the top rather than in line with the fraction bars}. Similarly, participants desired the ability to directly edit elements when placed (instead of using the sidebar panel to edit element attributes) and the ability to use shortcut keys for common actions such as copy, paste, and delete.

\textbf{\textit{Agent Training: User Control}} Participants also desired additional ways to guide agent learning, such as specifying what tutor fields the agent should be using to generate explanations for demonstrations it observed. P6 states: \emph{I might like to specify which of the input boxes I'm using to make this [new action]. Just to provide a little bit more structure on where exactly I'm getting the information from}. Similarly P13 said: "\textit{I wish I could've just [told] the thing...just multiple these two numbers...}". 

\textbf{\textit{Human-Agent Interactions}} P6, 11, 13, and 16 all expressed interest in additional interaction modalities for agent training such as providing additional input or being able to use verbal speech instead of typing. P16 said: "\textit{It [would] be cool to aside from like typing...saying it verbally, and then it automatically translates to what you're saying}".  Although participants did desire additional ways to interact with the agent, some also mentioned that the interactions felt "natural" as if teaching a student. For example, P18 and P16 said "\textit{It's essentially like coaching one student}" and "\textit{I felt like I was talking to a student.}", respectively.  Additionally, P7 noted "\textit{I think it was very natural. I felt like there was a character sitting somewhere and fixing my problem right away}". Participants did however wish that some aspects of agent training were less repetitive, such as having to label each step of the problem to generate task labels for the agent or the number of examples required for the agent to learn to solve the problem.

\textbf{\textit{Model Correctness}} Continuing from the previous theme, knowing when to terminate agent training was a question for many participants. Some believed firmly that the agents learned the tasks with P17 and P13, stating "\textit{I think the problems we did...it did it very well, and it learned the steps within 3 examples}", and "\textit{I really liked how it only took like 3 or 4 tries to really get the AI to really understand. It's obviously very smart}", respectively. However, others were not as sure. P12 said: "\textit{I didn't really have an indication as to how many times I would need to train it like maybe there could be a suggestion of like, do 5 or do 4?}". Likewise, when asked if they believed the agent fully learned the tasks P10 said "\textit{Possibly. I think I'd have to do it a little more to know}". 

\textbf{\textit{Tutorials \& User Guidance}} Participants stated that the tutorial component of the study was very beneficial as a primer to the system. P12 said "\textit{I found it pretty easy to use. But yeah, I think it would require like either a video tutorial at first...If you were going to utilize it for teachers}".  P10 expressed that the tutorial was necessary for them to adopt the system: "\textit{...at least the first time, if I didn't have [the video] to help me, I'd be like, this is too hard. I'm not using this.}" Additionally, others floated the idea of interactive guidance that allows them to step through a guided problem with P6 saying "\textit{Problems that I experienced were not with the agent, but with me, forgetting to fill out parts of the problem or hitting done too soon...I wish I had more step by step guidance for this}". Finally, P3 desired a tutorial that is always available for the user: "\textit{if there is a sample question or a sample way of teaching the agent on the side, it's gonna be very, very easy}". 

\textbf{\textit{Accessibility Features}} Some participants requested accessibility features such as larger text and text-to-speech functionality for visually challenged individuals. P17 said "\textit{I would have preferred a little bit bolder [text] if possible...I have little weak eyes, I'm sure many teachers would have little bit weaker eyes}". Additionally, P16 asked if they could "hear the message" referring to the agent's dialog messages instead of having to read them.  

\section{Discussion}


The results of our study are promising and indicate that teachers with varying degrees of technical experience can successfully use ATB to create and customize tutor interfaces and author expert models. 
First, all study participants were able to recreate both the Fraction Arithmetic (FR) and Square 25 (S25) tutors using the interface builder.  For the FR tutor, both the Low and High groups had comparable interface authoring times. The two groups, however, differed in interface authoring time for the S25 tutor, with the Low group taking approximately twice as long to author. Further investigation of the user logs showed that the Low group had a higher incidence rate of "delete" actions compared to the High group (Avg: 7.1 vs. 3.4). A "delete" action occurs when a user deletes an interface element. The difference in interface authoring time between the two groups can then be explained by the fact that the current tool does not let users reorder elements once placed. If an element is missed during interface creation, the user must backtrack and delete all elements placed after the point the initial element was missed. This directly contributed to the increased average authoring time for the Low group. These results indicate a need for a more flexible interface builder.  This is supported by multiple participants requesting the ability to edit interface elements directly (instead of using a separate editing console), reorder elements once placed, and adjust styling elements such as alignment.

We observed that the High group initially spent more time training their agents for the FR task, but subsequently spent less time on agent training in the S25 task. Our hypothesis was that those with more programming experience may initially be more cautious when training, but may more quickly develop an intuition for when the agent is sufficiently trained thus producing similarly accurate models in less time. To confirm any significant differences in the resulting expert models between both groups, we tested each model for accuracy. As presented in the previous section, all models between both groups achieved 100\% correctness except for one in the Low group. Upon examining further, it was noted that the corresponding participant only trained the agent in fraction multiplication using problems with at least one "1" in the numerators. Therefore, this agent only learned the edge case scenario where if one of the numerators contains a "1", it can just copy the other numerator to the answer field. This does not generalize to problems where neither of the numerators contains a "1".

Although most participants were able to author an accurate expert model, these findings indicate that we must provide a mechanism for testing learned expert models. Currently, users can gauge model accuracy by how many individual steps the agent solves correctly, but this does not provide a holistic view of the agent's knowledge. \cite{maclellan2022domain} suggest tracking a model's performance during training and terminating once an "acceptable level of performance" has been achieved. Taking this further, we might provide an indication of the model's performance to users to let them gauge when to terminate training. 

Overall, users see the value in ATB and have identified a set of potential benefits for both teachers and students. Study participants noted the potential time-saving benefits, evidenced by short tutor building and authoring times. Additionally, as presented in the affinity diagramming analysis, some participants expressed interest in integrating ATB into their classrooms. For students, participants identified benefits such as increasing the "teacher presence" and providing 24/7 help and practice when needed by the student. Additionally, one participant even suggested the tutor authoring process as a pedagogical tool. We can see that there is a genuine interest among teachers for a tool like ATB that provides the capability to create and customize tutor content for their classrooms. 

Finally, from our analysis, we identified a set of design recommendations for systems like ATB that researchers can use when developing similar systems.  

\textbf{\textit{Expected Features}} Users bring prior experience of  similar tools to the current system. As such, there must be a degree of familiarity with the basic functions of the system. Indeed, in our study, participants expressed the desire for more control over element order and alignment and the ability to directly edit elements instead of using the side panel edit component. These capabilities are quite common in many current interface builders. Deviating from established norms may hinder adoption. 

\textbf{\textit{Tutorials and Guidance}} Following the previous recommendation, users often require some sort of tutorial when acclimating to a new system. Multiple participants in our study expressed that the tutorial was a key component for understanding the system. Additionally, users requested real-time or quick-reference guidance that is always available for each component of ATB.  

\textbf{\textit{Flexible Design}} A flexible design for ATB involves giving users the \emph{option} to use multiple features, rather than shoe-boxing them into one particular design. Further, we want to provide users with the most flexible configuration possible while maintaining effectiveness of the system. For example, instead of pre-determining how the user interacts with the AI agent, we might give the user a set of ways to interact with the agent and leave the choice to them. In this case, we must, however, consider the ways in which the presence or absence of interactions, and thus the knowledge obtained from them, affects the agent's ability to learn. \cite{zhu2015machine} discusses this idea as an optimization problem where we want to maximize user satisfaction and experience while mitigating risk. We can apply this framework by ensuring that user needs are adequately addressed, while simultaneously ensuring all components of ATB are fail-proof.

\textbf{\textit{Testing for model correctness}} As we have explored, knowing when a model has achieved 100\% correctness is difficult task. We have seen in our study that some users are not fully confident that the agent has fully and correctly learned the problem when terminating training. We must therefore provide a way to test the learned expert models in real-time for users such as was suggested in \cite{maclellan2022domain}.  This will help users author robust expert models in the shortest time and using the least number of examples.

\section{Conclusions and Future Work}
To summarize, we presented the ATB system, which allows users to create and personalize tutors to fit their instructional needs. We analyzed the overall effectiveness and usability of the system by conducting a user study where 14 participants built tutors and trained expert models in Fraction Arithmetic and the Square 25 procedure. The findings of this study were instrumental in informing the next iterations of ATB.

We primarily wanted to investigate the ability for teachers with little to no specialized training in tutor authoring tools to use our system. By grouping our participants into those with little to no programming experience vs. those with moderate to high programming experience (the Low group and the High group), we were not only able to study how those in the Low group use our system, but also discover any latent differences between the two groups that could challenge any assumptions or biases we included in the design as researchers.

Informed by our study results, we plan to explore continued development of the interface builder to enable more flexible and intuitive design. This include the ability to reorder, edit, and align elements directly. Additionally, we plan to investigate new ways to interact with the AI agents. We plan to integrate language instruction for agent training, as well as the ability to guide the agent by selecting interface elements it should use in its explanation process.
In conclusion, this study is one of the first to evaluate interactive machine-learning based tutor authoring tools with teachers. We believe our work demonstrates the potential for teachers to use these tools to create and personalize intelligent tutors for their classes. We hope future work will further explore this potential.

\bibliographystyle{ACM-Reference-Format}
\bibliography{ATB-LAS-2024-BIB}

\end{document}